\documentclass[num-refs,a4paper, 12pt]{notre-template}
\usepackage{graphicx}
\usepackage{dcolumn}
\usepackage{bm}
\usepackage[utf8]{inputenc}
\usepackage[T1]{fontenc}
\usepackage{mathptmx}
\usepackage{latexsym}
\usepackage{hyperref}
\usepackage{academicons}
\usepackage[format=hang]{caption}

\usepackage[labelfont=bf,textfont=md]{caption}

\newcommand{\orcidlink}[1]{}

\usepackage{ulem}
\usepackage{xcolor}

\mathchardef\mhyphen="2D
\newcommand*\hBN{\ensuremath{\rm \textit{h} \mhyphen  BN}}
\newcommand*\refSI{S.I.}

\usepackage{adjustbox}
\usepackage{arydshln}
\usepackage{braket}
\usepackage{ulem}
\usepackage{tabu}
\usepackage{siunitx}
\usepackage[square,sort,comma,numbers]{natbib}
\usepackage{mathrsfs}
\usepackage{parskip}
\usepackage{mathtools,amssymb}

\begin{document}
\begin{center}
    \title{Evidence of Strong Dzyaloshinskii–Moriya interaction \\ 
at the cobalt/hexagonal boron nitride interface}
\end{center}

\maketitle

\author{Banan El-Kerdi\footnote{Banan.Kerdi@universite-paris-saclay.fr, Alexandra.Mougin@universite-paris-saclay.fr}~\orcidlink{0000-0002-6015-1850},}
\author{André Thiaville~\orcidlink{0000-0003-0140-9740},}
\author{Stanislas Rohart~\orcidlink{0000-0001-7772-1306},}
\author{Sujit Panigrahy~\orcidlink{0000-0002-8397-1325},}
\author{Nuno Brás,}
\author{João Sampaio~\orcidlink{0000-0002-0460-3664}},
\author{Alexandra Mougin\footnotemark[1]~\orcidlink{0000-0001-8607-0154}}
    \dedication{} 
\begin{affiliations}\\
Université Paris-Saclay, CNRS, Laboratoire de Physique des Solides, 91405 Orsay,  France
\end{affiliations}

\begin{abstract}
\centering\begin{minipage}{\dimexpr\paperwidth-4cm}


 \textbf{\Large{Abstract}} \\
 
 The Dzyaloshinskii–Moriya interaction (DMI) and perpendicular magnetic anisotropy (PMA) were measured on four series of Co films (1 to 2.2 nm thick) grown on Pt or Au and covered with h-BN or Cu. Clean h-BN/Co interfaces were obtained by exfoliating h-BN and transferring it onto the Co film in-situ in the ultra-high vacuum evaporation chamber. By comparing h-BN and Cu-covered samples, the DMI induced by Co/h-BN interface was extracted and found to be of comparable strength to Pt/Co, one of the largest known values. The strong observed DMI despite the weak spin-orbit interaction in h-BN supports a Rashba-like origin in agreement with recent theoretical results. By combining it with Pt/Co in Pt/Co/h-BN heterostructures, an even larger PMA and DMI is found which stabilises skyrmions at room temperature and low magnetic field.

\end{minipage}
\end{abstract}


Since the first exfoliation of graphene \cite{Novoselov2004}, the science of two-dimensional van der Waals (vdW) materials has revealed systems with radically new properties \cite{Cao2018,Wang2023}. 
Heterostructures based on vdW materials allow a great freedom of material stackings and interfaces with outstanding quality and multiple functionalities \cite{Geim2013}.
However, the interfaces between vdW materials and metallic thin films are still little explored, although they could also bring interesting results \cite{Wang2022}.
During the same period, new interfacial effects were observed in metallic magnetic heterostructures induced by strong spin-orbit coupling (SOC), leading to  Dzyaloshinskii-Moriya interaction (DMI) and perpendicular magnetic anisotropy (PMA), that stabilise chiral textures~\cite{thiaville2012dynamics,fert2013skyrmions,hellman2017}.
Among the new types of textures, skyrmions, which are easily moved by spin-orbit torques, form the basis for many new proposed spintronic devices~\cite{Sampaio2013,Fert2017}. This has motivated a search for systems with strong DMI.

The vdW/metallic ferromagnets could bring these two domains together.
A natural choice would be vdW with high SOC.
For example, a giant enhancement of PMA and current-induced spin-orbit torques by MoS$_2$~\cite{xie2019giant} and the valley splitting of WSe$_2$ and WS$_2$ monolayers on EuS \cite{zhao2017enhanced,norden2019giant} have been reported.
Yet, the observation of sizable DMI at graphene/Co interface~\cite{ajejas2018unraveling} or the current-induced magnetic switching of SrRuO$_3$/hexagonal Boron nitride (\hBN{})~\cite{xie2022rashba}  show that light 2D materials can also produce significant effects. 
Additionally, different spin filtering behaviors have been measured in Co/\hBN{}/Co and Co/\hBN{}/Fe tunnel junctions~\cite{Seneor2018}, which shows the diverse functionality of heterostructures based on \hBN.
This is striking given that \hBN{} is an insulator with negligible SOC~\cite{watanabe2004}, more often used as an encapsulation layer than a functional material in vdW heterostructures.

In this work, we focus on the Co/\hBN{} interface for which a large DMI and PMA were recently theoretically predicted~\cite{hallal2021rashba}.  
This prediction raises important questions concerning the origin and role of SOC in DMI and PMA. 
Additionally, density functional theory calculations predict that the Co work function  is modified by the chemisorption of \hBN{}, leading  to the formation of a large interfacial potential step~\cite{bokdam2014schottky}.  
A recent experimental study has shown, however, almost no DMI for Co deposited directly on \hBN{} \cite{rastikian2021magnetic}. Here, we developed a method to deposit \hBN{} flakes on clean Co layers in ultra-high vacuum (UHV). Using Brillouin light scattering (BLS) spectroscopy, we unambiguously show that this induces a sizeable DMI and PMA, with a magnitude comparable to the Pt/Co interface, which is one of the interfaces with the largest effects known to date. 
This leads to the formation of skyrmions at moderate applied fields in Pt/Co/\hBN{} heterostructures.

\section*{Experiments}

\subsection*{Sample preparation and measurements}

We prepared four series of samples containing  either a Ta/Pt or Ta/Au buffer layer, a Co ultra-thin film ($t_{\rm Co}$ ranging from 1 to 2.2~nm), and either a \hBN{} or metallic Cu/Al cap
(see further details in the Experimental Section and in the \refSI).
The metallic films were deposited by electron-beam evaporation in UHV.
The layers were grown on a Ta buffer to promote a (111) crystallographic orientation and Co layers of the same lattice constant as \hBN{} \cite{hallal2021rashba}.
The metallic cap was chosen to produce reference interfaces with low DMI~\cite{Ajejas2022}.

\begin{figure}[h]
\centering
  \begin{center} \includegraphics[width=0.8\textwidth]{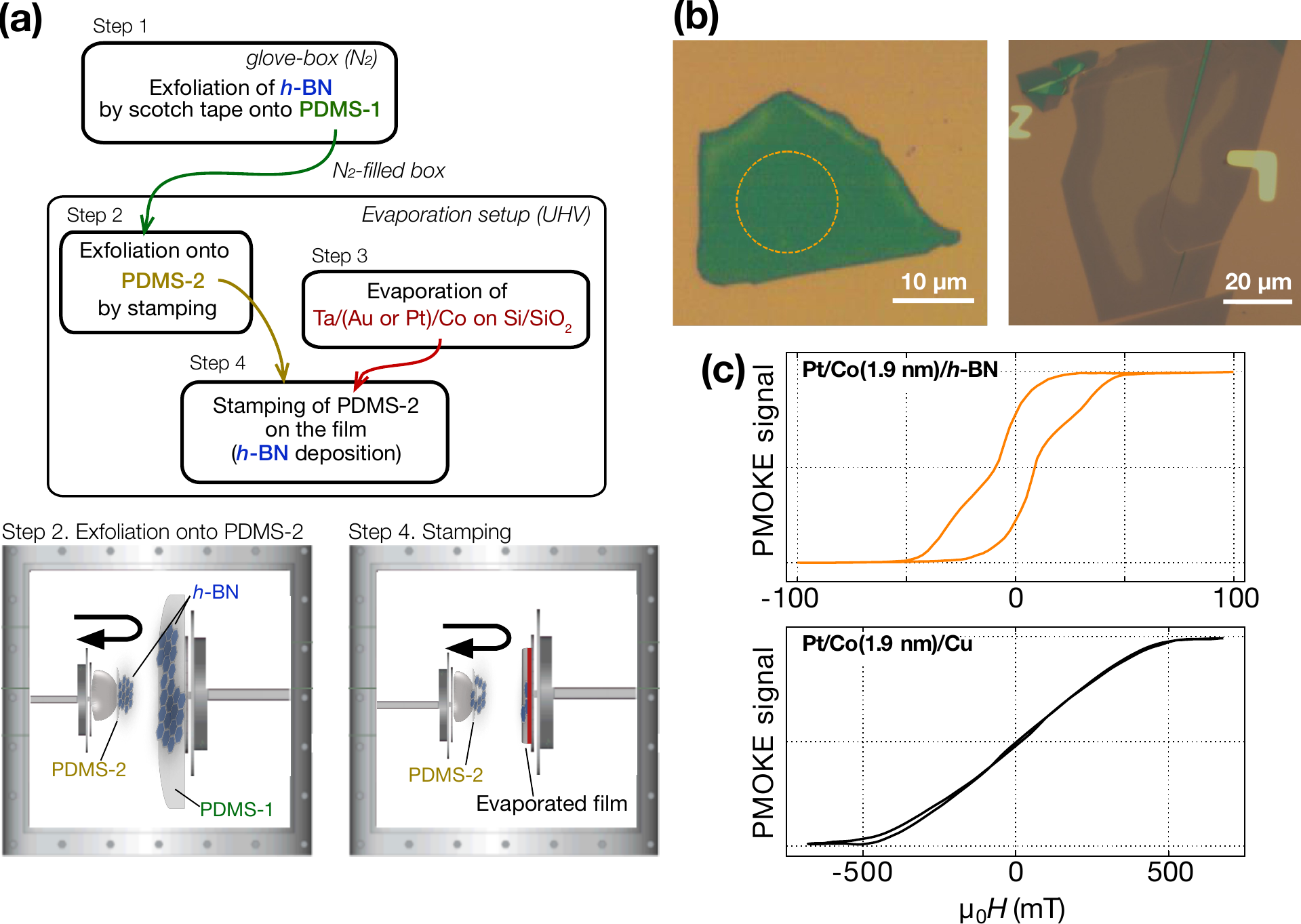}  \end{center}
  \caption{ \textbf{(a)} Schematic of the fabrication steps of a sample with a Co/\hBN{} interface and schematic drawings of the mechanical exfoliation in UHV. \textbf{(b)}  Micrographs of the studied \hBN{} flakes on cobalt thin films. The lateral size of the flake is $\sim${30} $\mu$m and the thickness is $\sim$ {40 nm}. \textbf{(c)}   Hysteresis cycle measured by PMOKE of a Ta/Pt/Co (1.9 nm)/\hBN{} film covered by \hBN{} in the region shown by the circle in b)  (orange) and on a control sample made with Ta/Pt/Co (1.9 nm)/Cu/Al (black). The signals were corrected to remove a linear contribution from the Faraday effect.}.  
  \label{fig:BLS}
\end{figure}

For the series capped with \hBN{}, an exfoliated \hBN{} flake was then transferred to the Co film by stamping a PDMS adhesive, onto which \hBN{} flakes had been previously exfoliated (see \textbf{Figure~\ref{fig:BLS}}a and Experimental Section). The final \hBN{}  exfoliation and transfer onto the Co film were done in-situ, in UHV to prevent Co oxidation. 
The flakes are typically a few tens of µm in width and tens of nm in thickness, as confirmed by atomic force microscopy (see Supporting Information).
\textbf{Figure~\ref{fig:BLS}}b shows optical microscopy images of typical \hBN{} exfoliated flakes deposited on Co (more examples are given in the Supporting Information.
 The difference in color of the flakes is due to their different thickness. In many flakes, a difference in optical contrast between the center and the border is visible, as in the right-hand flake in \textbf{Figure~\ref{fig:BLS}}b. 

The sample magnetic properties were measured at room temperature with polar magneto-optical Kerr microscopy (PMOKE), which is sensitive to out-of-plane magnetization. 
\textbf{Figure~\ref{fig:BLS}}{c} shows the hysteresis loops of the Pt/Co (1.9 nm)/\hBN{} sample (on the flake region circled in \textbf{Figure~\ref{fig:BLS}}{b}) and of the Pt/Co (1.9 nm)/Cu control sample.
The Pt/Co/\hBN{} film has a perpendicular magnetic anisotropy evidenced by the partial out-of-plane remanence, contrary to the control sample. 
For all sample series, thinner Co films are perpendicularly magnetized while thicker Co films are magnetized in-plane.
This reveals a competition between a perpendicular interfacial anisotropy $K_s$ (with $K_s$ in J/m$^2$) and the dipolar-induced shape anisotropy, as described by the anisotropy field
\begin{equation}
     H_K =  \frac{2 K_s}{  \mu_0 M_s t_{\rm Co}} -M_s 
     \label{eq:Hk}
\end{equation}
where $M_S$ is the spontaneous magnetization and $\mu_0$ is the vacuum permeability. At small thicknesses, the interfacial anisotropy dominates and the magnetization is perpendicular ($H_K>0$). The reorientation thickness (where $H_K=0$) depends on $K_s$, therefore on the anisotropies induced by the buffer and capping layers.

\begin{figure}[!h]
    \centering
  \includegraphics[trim=2 0 5 1,clip,width=1\textwidth]{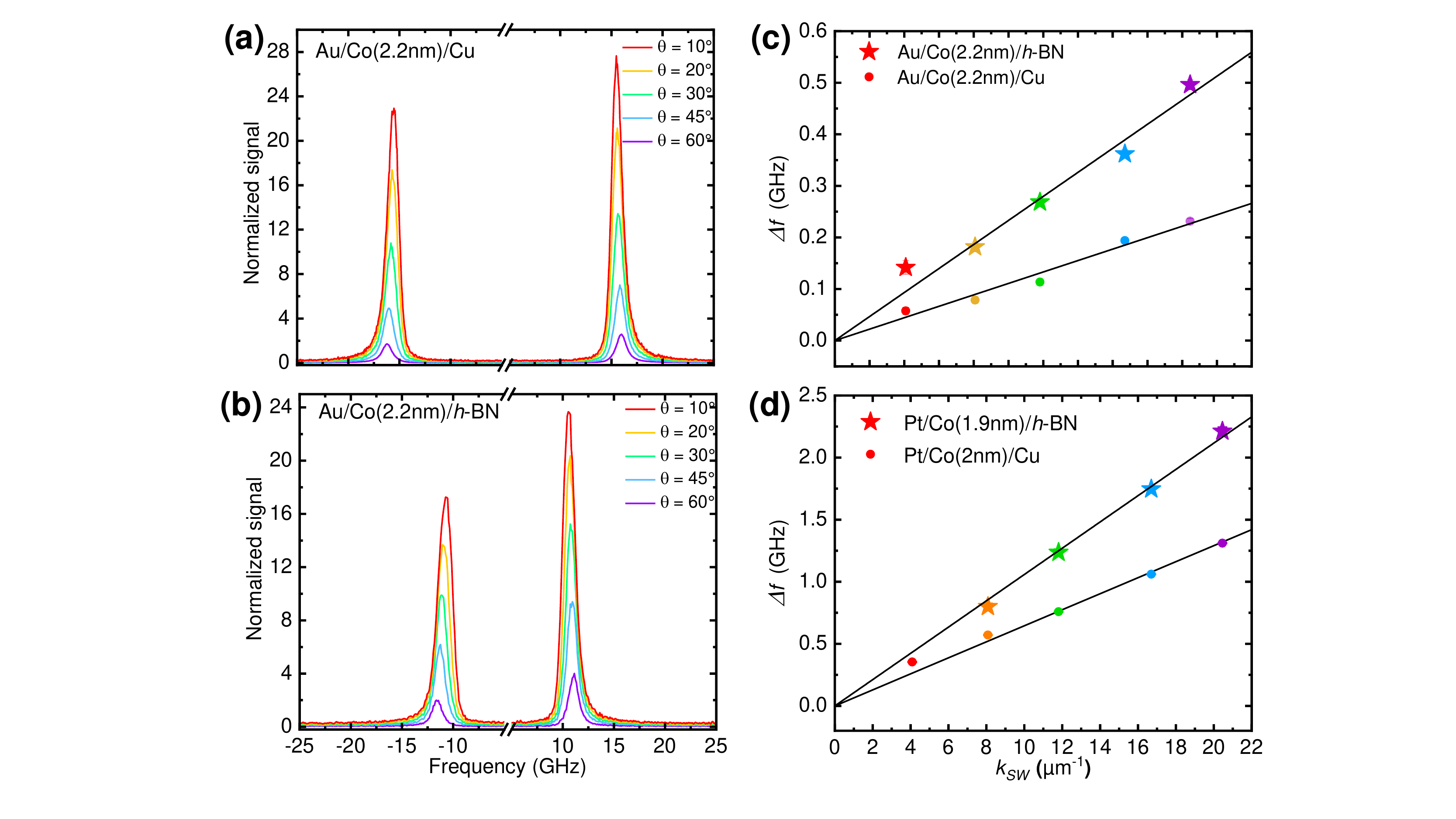}
  \caption{ 
  (\textbf{a,b}) BLS spectra at different incidence angles $\theta$ for \textbf{(a)} Ta/Au/Co(2.2nm)/Cu ($H_{\rm ext}=100$~mT) and \textbf{(b)}  Ta/Au/Co(2.2 nm)/\hBN{} ($H_{\rm ext}=148$~mT). 
  \textbf{(c,d)} S-AS frequency shift ($\Delta f$) as a function of the spin-wave wavevector ($k_{\rm SW}$) for \textbf{(c)} Au/Co(2.2 nm)/Cu and Au/Co(2.2 nm)/\hBN{} and for \textbf{(d)}  Pt/Co(2 nm)/Cu and Pt/Co(1.9 nm)/\hBN{}. The circles and stars correspond to the experimental data and the line is a linear fit used in the determination of the DMI strength. The error bars (often smaller than the symbols) show the error from the fitting of the peaks.}
  \label{fig:deltaf}
\end{figure}

We use Brillouin light scattering (BLS) spectroscopy to extract locally the strength of DMI and PMA by analyzing the frequency shift of backscattered photons from an impinging laser beam.
The measurements are carried out in the Damon-Eshbach configuration \cite{damon1961magnetostatic} with an external magnetic field $H_{\rm ext}$ that saturates the magnetization along an in-plane direction. 
An aperture allows to probe locally the properties on the flake over a diameter of 15 to $30~\mu$m. 
 The spectra consist of two peaks, the Stokes (S) and the anti-Stokes (AS), associated with counter-propagating spin waves (SWs), which are affected by DMI in opposite ways, and so present a frequency difference given by~\cite{belmeguenai2015interfacial,schlitz2021control}
 \begin{equation}
     \Delta f(k_{\rm SW}) = |f_{AS}| - |f_{S}| =  \frac{2 \gamma}{\pi M_s} \frac{D_s}{t_{\rm Co}} k_{\rm SW},
     \label{eq:DF}
  \end{equation}
where $k_{\rm SW}$ is the spinwave wavector, $\gamma$ is the gyromagnetic ratio, and $D_s$ quantifies the DMI interfacial strength and sign. 
The quality of the spectra allows for the detection of frequency differences down to a few tens of MHz. 
The incidence angle $\theta$ is varied between 10° and 60° to vary the SW wavevector, as $k_{\rm SW} = 4 \pi \sin (\theta)/\lambda$, where $\lambda= 532$~nm is the laser wavelength.  
We assumed $M_s = 1.15\pm0.05$~MA/m for all samples based on magnetometry measurements of Cu-capped samples (see Experimental Section). 
Experimental reports on these metallic interfaces~\cite {lau2012,Gweon2018,Wilhelm2004} and band structure calculations of the Co/\hBN~interface show that the magnetic moment is not modified so the magnetization is considered constant across the cobalt thickness.
Likewise, the gyromagnetic ratio was assumed to be the same for all samples~\cite{shaw2013,shaw2014,Shaw2021} and was measured on a Au/Co/\hBN{} sample (see Supporting Informtion).

\textbf{Figures~\ref{fig:deltaf}}(a,b) shows BLS spectra for different wave vectors for two Au/Co based samples, covered by Cu and \hBN{}. 
A small difference between S and AS peak frequencies ($\Delta f$) is seen, indicating an asymmetric dispersion $f(k)$ induced by the DMI (Equation~\ref{eq:DF}). 
Fitting the peaks (see Supporting Information), $\Delta f$ is evaluated and plotted as a function of $k_{\rm SW}$ (\textbf{Figure~\ref{fig:deltaf}}c).
For both samples, $\Delta f$ increases linearly with $k_{\rm SW}$, as expected for a film with DMI  \cite{moon2013spin,belmeguenai2015interfacial} and $D_s$ can be extracted using Equation~\ref{eq:DF}. 

We emphasize that the DMI sign can also be determined (see the Experimental Section). 
In all measurements, the $D_s$ is negative (using the sign convention of ref.~\cite{belmeguenai2015interfacial}), so left-handed cycloids are favored.

In the limit of $k_{\rm SW}\approx0$, the mean of the  Stokes and anti-Stokes frequencies, $f_0$, is given by \cite{kittel1948theory}
\begin{equation}
    2 \pi f_0 = \mu_0 \gamma \sqrt{H_{\rm ext} (H_{\rm ext} - H_K)} ,
    \label{eq:Kittel}
\end{equation}
Although $k_{\rm SW}\neq 0$ in BLS measurements and additional contributions to $f_0$ are present~\cite{Kalinikos1986}, these remain small and were neglected (see Supporting Information).
The interfacial anisotropy parameter $K_S$ can then be deduced from $f_0$ using Equations~\ref{eq:Hk} and \ref{eq:Kittel}.

\subsection*{Results: thickness dependence of the interfacial DMI and PMA}

\begin{figure}[!h]
\centering
\includegraphics[trim= 70 0 1 0,clip,width= 1.1\textwidth]{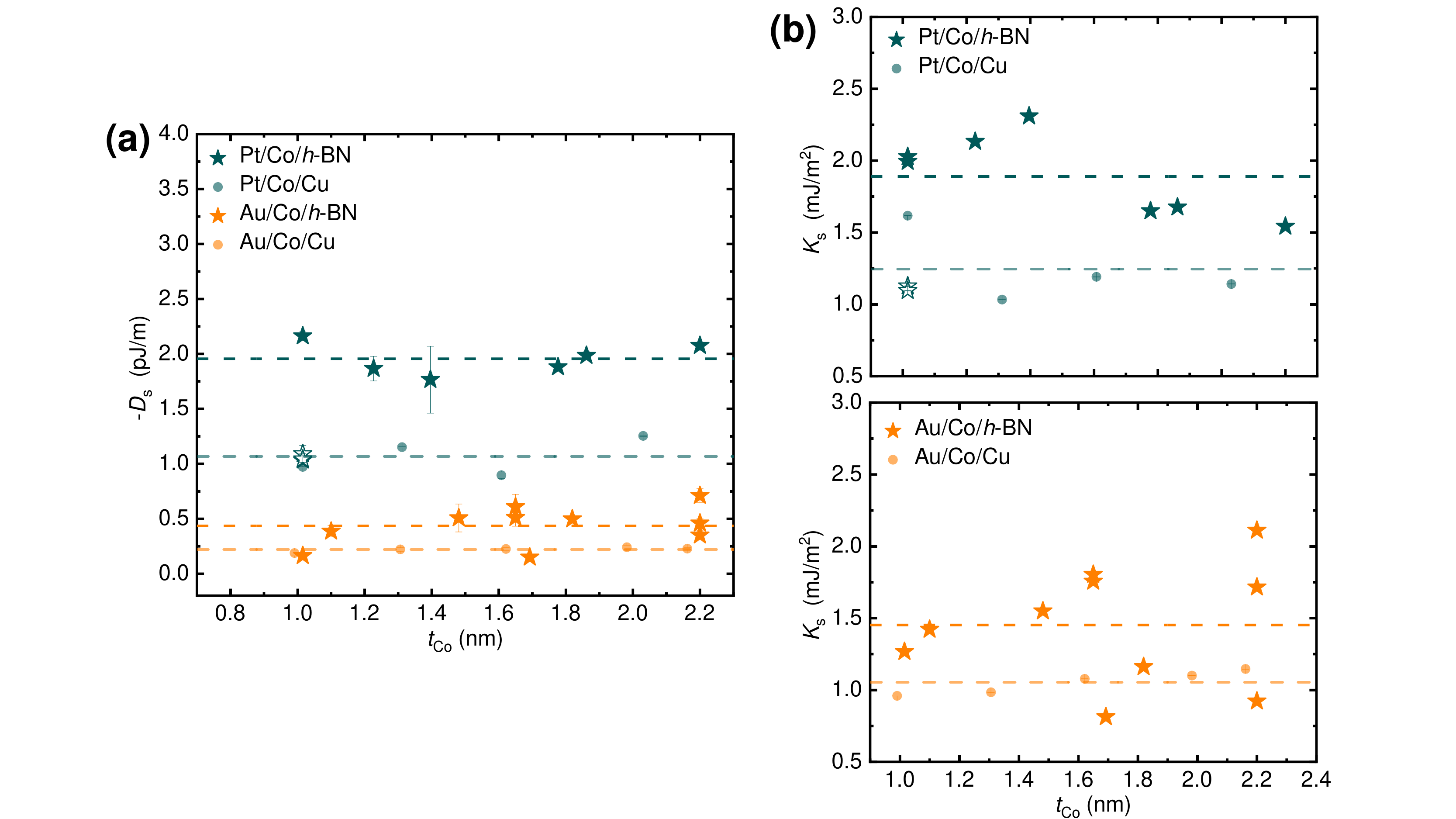}
\caption{
\textbf{(a)} Interfacial DMI strength $D_s$ and \textbf{(b)} interfacial anisotropy strength $K_s$ versus Co thickness for Pt/Co/\hBN{},  Pt/Co/Cu, Au/Co/\hBN{} and Au/Co/Cu. 
The dashed lines indicate the average values. The empty stars in (a) are measurements of the Pt/Co/\hBN{} series that were not included in the averages. Note that some error bars (estimated from the fits) are smaller than the symbols. 
}
\label{fig:DsandKs}
\end{figure}

To extract the \hBN{} contribution to DMI and PMA, the four sample series were compared as shown in \textbf{Figure~\ref{fig:DsandKs}}.
The values of $D_s$ vary only marginally with the Co layer thickness, confirming the interfacial nature of the DMI.
The $D_s$  values are more dispersed for  \hBN{}-capped samples.
In addition to the error due to the fitting itself, this dispersion might be due to the flake-to-flake variability on a given sample (see \refSI{}). 
Even under some single flakes, the Co/\hBN{} interface may not be homogeneous as suggested by different optical contrast between the center and the border. 
Such a local variation is hard to probe given the spot size ($>15$~µm) of our BLS setup.

\begin{table*}[h]
\centering
\begin{tabular}{ccc}
                        & $D_s$ (pJ/m)          & $K_s$ (mJ/m$^2$)  \\ \hline
 \textbf{Pt/Co/Cu}      & $-1.07 \pm 0.16$      & $1.25 \pm 0.30$   \\ 
 Co/Cu                  & $\approx 0$           & 0.2               \\ 
 Pt/Co                  & $-1.07 \pm 0.16$      & $1.05 \pm 0.30$   \\ \hline

 \textbf{Pt/Co/\hBN{}}  & $-1.96 \pm 0.15 $     & $1.89 \pm 0.30$   \\ 
 Co/\hBN{}              & $-0.89 \pm 0.22$      & $0.84 \pm 0.42$   \\ \hline
 \textbf{Au/Co/Cu}      & $-0.22 \pm 0.02$      & $1.05 \pm 0.08$   \\ 
 Au/Co                  & $-0.22 \pm 0.02$      & $0.85 \pm 0.08$   \\ \hline
 \textbf{Au/Co/\hBN{}}  & $-0.43 \pm 0.18$      & $1.45 \pm 0.41$   \\ 
 Co/\hBN{}              & $-0.21 \pm 0.18$      & $0.60 \pm 0.41$   \\ \hline
\end{tabular}
\caption{\label{tab:table1}  Mean values of $D_s$ and $K_s$ for the four sample series (from \textbf{Figure}~\ref{fig:DsandKs}). The error estimations are the standard deviation of the data points. For each series, the contribution of each interface is deduced, assuming 
$K_s^{Co/Cu} =0.2$~mJ/m$^2$\cite{Johnson1996} and $D_s^{Co/Cu} = 0$~pJ/m\cite{Ajejas2022}. The values found for the Pt/Co and Au/Co are in good agreement with literature~\cite{yang2015,zhu2022}.}
\end{table*}

\textbf{Table}~\ref{tab:table1} summarizes the measured values of $D_s$ for the four sample series.  
As the magnetic layers are thick enough~\cite{BlancoRey2022}, DMI can be considered as an additive effect, where the $D_s$ of a X/Co/Y stack can be decomposed as the sum of its two interfaces: $D_s^{X/Co/Y}$  = $D_s^{X/Co} + D_s^{Co/Y}$. Note that the stacking order has to be taken into account since $D_s^{X/Co} =  - D_s^{Co/X}$. 
Experimentally, it is not possible to separate the two interfacial contributions, and a comparison with a reference sample is required.
As Cu is a light element and with a work function close to that of Co, we expect a very small Co/Cu contribution~\cite{fert1980,Ajejas2022}  which is therefore neglected.
We thus find  $D_s^\mathrm{Pt/Co} = -1.07$~pJ/m.
Knowing this value, the contribution of Co/\hBN{} interface is deduced from the Pt/Co/\hBN{} series as $D_s^\mathrm{Co/\hBN{}} = -0.89$~pJ/m, a very large value.
The same reasoning can be used for the two series with the Au buffer, resulting in a lower value $D_s^\mathrm{Co/\hBN{}} = -0.21$~pJ/m,
a difference that may be attributed to buffer-dependent structural~\cite{voigtlander1991,lundgren2000,baudot2004} and electronic~\cite{kyuno1996theoretical, bandiera2012enhancement} properties of the Co film.

The $K_s$ parameter does not show any significant variation with Co layer thickness (\textbf{Figure}~\ref{fig:DsandKs}b) as well, confirming its interfacial origin. As for $D_s$, $K_s$ is the sum of the two interface contributions and the same arithmetic as above can be used, considering $K_s^{\rm Co/Cu}=0.2$~mJ/m$^2$ \cite{Johnson1996}. 
We find that the interfacial anisotropy of Co/\hBN{} interface, $K_s^\mathrm{Co/\hBN{}}$, is $0.84 $~mJ/m$^2$ for the Pt stacks and $0.60 $~mJ/m$^2$ for the Au stacks.
The difference with the buffer and the dispersion in the values for the Co/\hBN{} samples may be explained with the same arguments as for $D_s$.

Remarkably, Pt/Co and Co/\hBN{} induce a DMI of the same sign.
Films with large DMI typically contain one Pt/Co interface, which produces one of the largest known values of $D_s$~\cite{belmeguenai2015interfacial, Kuepferling2020}. As reversing the interface reverses the sign of the DMI (i.e., one expects $D_s^{\rm Pt/Co} \approx - D_s^{\rm Co/Pt}$), it is not possible to double the DMI by simply adding a second Pt layer (e.g., Pt/Co/Pt). 
Our results pave the way to an increased DMI, since in Pt/Co/\hBN{} the contributions of the two interfaces are significant and add up.

\subsection*{Magnetic skyrmions}

\begin{figure}[!h]
  \includegraphics[trim= 1 220 50 100,clip,width=1\textwidth]{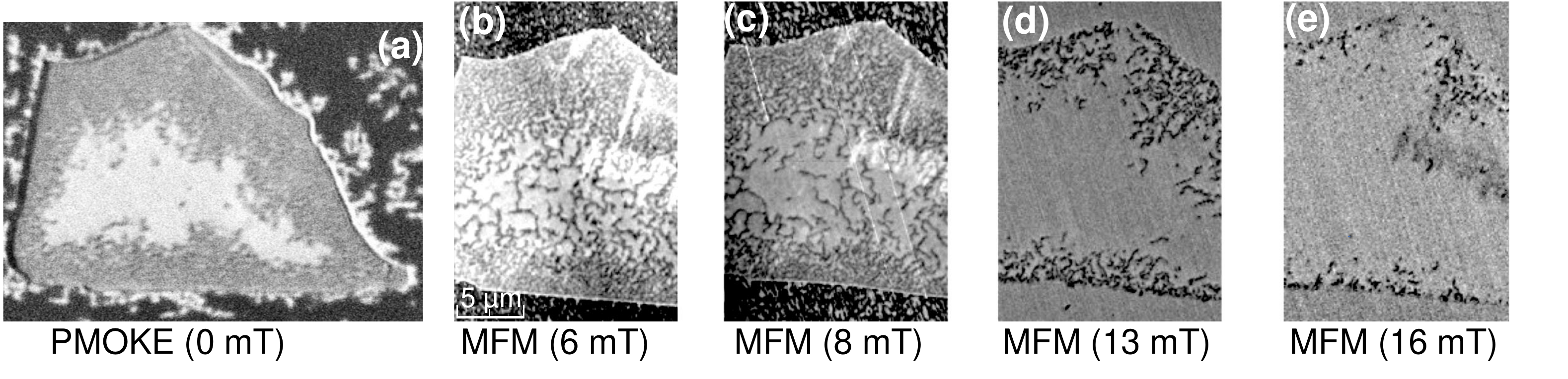}
  \caption{ Magnetic textures in the Pt/Co(1.9 nm)/\hBN{} sample imaged by \textbf{(a)}  PMOKE at zero magnetic field, and by \textbf{(b-e)} magnetic force microscopy (MFM) at different fields.} 
  \label{fig:MFM}
\end{figure}

The large contribution of the Co/\hBN{} interface to DMI should improve the stability of chiral magnetic textures at room temperature.
The hysteresis cycle shown in \textbf{Figure\ref{fig:BLS}c} for Pt/Co (1.9 nm)/\hBN{} is not saturated at low field, which is a fingerprint of magnetic textures including skyrmions.
\textbf{Figure~\ref{fig:MFM}} shows images of the magnetic textures at different applied fields. Striped, labyrinthine domains are observed at low field, with a typical width of  $\approx$ 300~nm (measured in the center of the flake). When a field is applied, these stripes collapse and very small bubble-like domains appear, of a typical size of $\approx$ 140 nm. Bubble-like domains are likely to be skyrmions as the measured DMI is large enough to impose a well-defined chirality \cite{thiaville2012dynamics} and, thus, topology \cite{fert2013skyrmions}.

A difference is found between the center of the flake and its periphery, with larger domains observed in the center of \textbf{Figure~\ref{fig:MFM}}b. Larger domains may be caused by lower magnetization or higher domain wall energy (or both)~\cite{Kooy1960,Rohart2013}. 
Note that even slight variations of any magnetic parameter (magnetization, anisotropy or DMI that may be different at the center) change significantly the size of stripes or skyrmions.

Theory and rare experiments state that the local electronic structure of the Co/\hBN{} interface drives its magnetic properties. A host of factors enable the interface engineering of such heterostructures (crystallography, presence of local structural or chemical defects, hybridization, oxidation, locally embedded charges…). In the theoretical study based on electronic structure calculations of ref.~\cite{hallal2021rashba}, a key point was that Co and \hBN{} formed a hexagonal lattice with a tiny mismatch (0.12\%). The calculations found that the AB stacking was slightly lower in energy than the AC stacking with differences of 10\% in PMA and 20\% in DMI. In our samples, even if perfect \hBN{} crystals were deposited on the Co layer, such ideal stackings are unlikely over a significant area fraction  because the Pt layer consists of 15-nm-sized (111)-textured grains oriented randomly in-plane (see \refSI{}) and the \hBN{} lattice undergoes smooth undulations of crystalline orientation over a flake. 
In such conditions, the DMI value reported here, about one-half of the calculated one, suggests a more general cause. 
In a recent analogue  study~\cite{xie2022rashba}  of \hBN{}/SrRuO$_3$ interface with an enormous lattice mismatch (30\%), the crystallographic incompatibility was addressed in an effective tight-binding model that bears no relation to the lattices of the two compounds, and in which the key parameter is the hybridization between the \hBN{} $p_z$ and the Ru 3d orbitals. 
Thus, the high DMI can mostly be attributed to hybridization at the interface, relying only on the Co/\hBN{} proximity, a feature which is expected to depend much less on a precise crystallographic structure. Given the observed chemisorption~\cite{bokdam2014schottky}, the  work function of Co interfaced with \hBN{} should decrease regardless of the detailed crystallography. This supports a DMI physical origin different from the one observed at heavy metal/ferromagnetic material interfaces, where DMI originates from a superexchange through non-magnetic atoms~\cite{fert1980,Fert1990,yang2015} and is sensitive to atomic stacking. Instead for Co/\hBN, the Rashba-induced DMI only requires a hybridization at the interfaces~\cite{yang2018,hallal2021rashba,Yang2023}.

In conclusion, we have reported the experimental observation of large DMI and PMA induced by \hBN{} flakes on a Co thin film, in sharp contrast to the conventional use of \hBN{} as inert material to encapsulate other 2D van der Waals materials.
In addition to being relatively strong (comparable to the record values in metallic systems), the DMI  of Co/\hBN{} is of a sign that reinforces the DMI of Pt/Co when combined. Interestingly, this strong effect in the absence of high spin-orbit in \hBN{} supports the hybridization Rashba-like mechanism as a possible source of DMI.
Moreover, the observed DMI rivals the predicted one in calculations for epitaxial systems whereas our samples are not.
  
 Our work thus reveals a new functionality of \hBN{}  with consequences to  other branches of physics such as  electronic structure, topological defects and 2D magnetism, as illustrated by the observation of stable skyrmions at room temperature with a low magnetic field in Pt/Co/\hBN{}.

\section*{Experimental Section}

All measurements were performed at room temperature.

\threesubsection{Sample preparation}
Metallic layers are deposited by electron-beam evaporation in ultra-high vacuum (base pressure $10^{-10}$~mbar), on Si(001)/SiO$_2$(300 nm) substrates (the SiO$_2$ layer is meant to enhance magneto-optical signals~\cite{hrabec2017making}). 
 The thicknesses of the buffers are Ta~(2.7~nm)/ Pt~(4~nm) or Ta~(2.7~nm)/ Au~(4.5~nm), and that of the metallic cap are Cu~(3~nm)/ Al~(3~nm). 

Flakes of few-layer \hBN{} are mechanically exfoliated from a commercial \hBN{} crystal in a glove box using the scotch tape method. Then a first PDMS adhesive ('PDMS-1') is brought into contact to transfer the \hBN{} flakes on top of it (\textbf{Figure\ref{fig:BLS}}a, step 1).
Next, the PDMS-1 is transferred quickly to the UHV chamber using a box filled with N$_2$.
A second exfoliation with another PDMS adhesive ('PDMS-2') is done in UHV (step 2) to provide a clean \hBN{} surface.
Then, the \hBN{} is transferred onto the Co stack, which was previously deposited on a Si/SiO$_{2}$ substrate  (steps 3 and 4). 
Finally, the complete stack (Ta/Au/Co/\hBN{} or Ta/Pt/Co/\hBN{}) is annealed at 100°C in UHV.

 \threesubsection{Brillouin light scattering (BLS)}
 BLS experiments are performed in the  Damon–Eshbach configuration by illuminating the sample with a green laser (30 µm diameter) at variable incident angle $\theta$ (10° to 60°). The laser power is 30 mW to avoid sample overheating. The area of interest is selected by a 15 to 30~$\mu$m aperture located on the backscattered beam path.
 In our configuration, the transferred optical wavevector \textbf{$k_{\rm SW}$} (along \textbf{$\hat{x}$}), the applied field (along \textbf{$-\hat{y}$}) and the film normal (from the substrate to Co, \textbf{$\hat{z}$)} form a right-handed frame. In this geometry, $D_s$ has the opposite sign of $\Delta f$.
 The scattered light is collected by the focusing lens and its spectrum is analyzed by triple-pass tandem Fabry-Perrot TFP-2 HC spectrometer. The acquisition time  is chosen so to obtain at least 300 counts at the peak maximum. The shown spectra were normalized to the measurement time (and are shown as average counts per 10 scans of the spectrometer).

\threesubsection{Polar magneto-optic Kerr microscopy}
PMOKE measurements are performed using a commercial Evico/Zeiss microscope.
The image in \textbf{Figure~\ref{fig:MFM}}a  was obtained by subtracting a reference image taken  under a 100~mT saturating field.

\threesubsection{Magnetic force microscopy (MFM)}
MFM images are obtained using a two-pass technique using a Park Technologies NX10 microscope: a tapping-mode pass provides the topography, and a second pass, at a lift height of 20~nm, measures the magnetic signal from the phase variation of the tip oscillations. The tip is coated with CoCr/Cr, whose thickness was adjusted to enhance the magnetic contrast on the \hBN{} flake. Out of the \hBN{} flake, the tip stray field perturbs the magnetic textures, since it is much closer to the magnetic layer than when on top of the \hBN{} flake.

 \threesubsection{Magnetometry}
 The magnetization was measured by SQUID magnetometry on Au/ Co~(1nm)/ Cu and Pt/ Co~(1nm)/ Cu samples at room temperature. 
 The magnetization was evaluated by the difference at remanence, after positive and negative saturation with an out-of-plane magnetic field. We obtained $M_S = 1.1$~MA/m and $1.2$~MA/m for the Au and Pt based samples, respectively, in good agreement with literature~\cite{zhang1991pt, Gweon2018}. Considering a 10\% uncertainty inherent to the technique, these values are considered to be similar and $M_S = 1.15$~MA/m was used in the analysis.

\medskip
\textbf{Supporting Information} \par 
The Supporting Information is available free of charge on the
ACS Publications website at \\ 
https://pubs.acs.org/doi/10.1021/acs.nanolett.2c04985.

It includes X-ray diffraction measurements of the metallic heterostructures, atomic force microscopy on h-BN flakes, the fitting procedure of the BLS data, determination of the effective gyromagnetic factor g through BLS experiments performed at different magnetic fields, determination of the magnetic anisotropy as a function of the Co thickness for the 4 series of samples, summary table including all samples with all physical parameters.

\medskip
\textbf{Acknowledgements} \par
We acknowledge fruitful discussions with Jacques Miltat, Vincent Jeudy and Aurélien Manchon. We thank Catherine Journet at University Claude Bernard for providing \hBN{} crystals used in preliminary experiments and Antoine Gallo-Frantz for X-ray diffraction experiments in LPS. We thank the staff of the MPBT (physical properties - low temperatures) platform of Sorbonne Université for SQUID measurements.
This work was supported by the Agence Nationale de la Recherche (ANR) through the project ACAF (ANR-20-CE30-0027), and by the  public grant overseen by the ANR as part of the ‘Investissement d’Avenir’ programme (LABEX NanoSaclay, ANR-10-LABX-0035 SPiCY and ANR-10-LABX-0035 CHIVAWAA).



\providecommand{\noopsort}[1]{}\providecommand{\singleletter}[1]{#1}%

\end{document}